\documentclass[]{article}
\usepackage{graphicx}
\title{Rindler approximation to Kerr black hole}
\author{\bf H. A. Camargo $\cdot$ M. Socolovsky}
\date{}
\begin{document}
\makeatletter{\renewcommand*{\@makefnmark}{}
\maketitle

{\bf Abstract} We show that the Rindler approximation to the time-radial part of the Kerr and Kerr-Newman metrics near their external $h_+$ and internal $h_-$ horizons {\bf only} holds {\bf outside} $h_+$ and {\bf inside} $h_-$, so respectively inside and outside the external and internal ergospheres, regions where, in Boyer-Lindquist coordinates, both $g_{tt}$ and $g_{rr}$ are negative, but preserving the Lorentzian character of the metric, and $r>0$ i.e. outside the region $r<0$ where closed timelike curves exist. At each point, the choice of Rindler coordinates is not trivial, but depends on the polar angle $\theta$. The approximation, as is known, automatically gives the absolute values of the surface gravities $\kappa_\pm$ as the corresponding proper accelerations, and therefore the Hawking temperatures $\tau_\pm$ at $h_\pm$.


{\bf 1 Kerr-Newman black hole and Rindler space}

It is well known that near the horizons of the Schwarzschild, Reissner-Nordstrom, Kerr and Kerr-Newman black holes, the time-radial part of the geometry can be approximated by the 2-dimensional Rindler space-time, with the proper acceleration representing the absolute value of the corresponding surface gravities. It is interesting to investigate, mainly in the case of two horizons, on which side of the outer ($h_+$) and inner ($h_-$) horizons the approximation holds. This is done here for the Kerr and Kerr-Newman cases, where it is shown that it occurs within the domains of outer communication, that is, outside the black hole and white hole regions i.e. outside $h_+$ and inside $h_-$. The result is known and expected for the outer horizon, but not so for the inner horizon.

The Kerr-Newman [1-2] metric in Boyer-Lindquist [3] coordinates is given by $$ds^2={{\Delta}\over{\Sigma}}(dt-asin^2\theta d\varphi)^2-{{\Sigma}\over{\Delta}}dr^2-\Sigma d\theta^2-{{(r^2+a^2)^2}\over{\Sigma}}sin^2\theta(d\varphi-{{a}\over{r^2+a^2}}dt)^2,\eqno{(1)}$$ where $t,r\in(-\infty,+\infty)$, $\varphi\in[0,2\pi)$, $\theta\in[0,\pi]$, $\Delta=r^2+a^2-2Mr+Q^2$, $\Sigma=r^2+a^2cos^2\theta$, $M$ is the gravitational mass, $a$ the angular momentum/unit mass, and $Q^2=q^2+p^2$, with $q$ the electric charge and $p$ the magnetic (Dirac) charge. The Kerr-Newman metric reduces to the Kerr metric when $Q^2=0$. In the following we shall consider the case $M^2-(a^2+Q^2)>0$.

-------------------------------------

H. A. Camargo

Facultad de Ciencias, Universidad Nacional Aut\'onoma de M\'exico

Circuito Exterior, Ciudad Universitaria, 04510, M\'exico D.F., M\'exico

e-mail: hugo.camargo@correo.nucleares.unam.mx

M. Socolovsky

Instituto de Ciencias Nucleares, Universidad Nacional Aut\'onoma de M\'exico

Circuito Exterior, Ciudad Universitaria, 04510, M\'exico D.F., M\'exico

e-mail: socolovs@nucleares.unam.mx

Let us investigate this metric near the Killing horizons $h_+$ ({\it event} horizon) and $h_-$ ({\it Cauchy} horizon), respectively at the roots of $\Delta$: $$r_\pm=M\pm\sqrt{M^2-(a^2+Q^2)}.\eqno{(2)}$$ In the neighborhood of $r_\pm$, we define the radial coordinate $\rho$ through $$r:=r_\pm+{{\alpha_\pm}\over{r_\pm}}\rho^2 \eqno{(3)}$$ with $[\rho]=[L]^1$ and $\alpha_\pm$ functions to be determined later, with $[\alpha_\pm]=[L]^0$. We consider the different terms in (1) up to $O(\rho^2)$.

i) In the last term of (1) we have, up to $O(\rho^2)$, $$d\varphi-{{a}\over{r^2+a^2}}dt=d\varphi-{{adt}\over{(r_\pm+{{\alpha_\pm}\over{r_\pm}}\rho^2)^2+a^2}}=
d\varphi-{{adt}\over{r_\pm^2+a^2+2\alpha_\pm\rho^2}}$$ $$=d\varphi-{{adt}\over{r_\pm^2+a^2}}(1-{{2\alpha_\pm}\over{r_\pm^2+a^2}}\rho^2)=d\varphi-\omega_\pm dt+{{2a\alpha_\pm\rho^2}\over{(r_\pm^2+a^2)^2}}dt$$ $$=d\tilde{\varphi}_\pm+{{2a\alpha_\pm\rho^2}\over{(r_\pm^2+a^2)^2}}dt, \eqno{(4)}$$ where $$\omega_\pm=\omega(r_\pm,\theta)={{a}\over{r_\pm^2+a^2}}\eqno{(5)}$$ is the dragging angular velocity of spacetime due to the rotation of the hole [4] $$\omega(r,\theta)={{a(2Mr-Q^2)}\over{(r^2+a^2)^2-a^2sin^2\theta\Delta}}\eqno{(6)}$$ evaluated at the horizons $h_\pm$, and $$\tilde{\varphi}_\pm=\varphi-\omega_\pm t \eqno{(7)}$$ are co-rotating azimuthal angles. (To each horizon one can associate a co-rotating coordinate system, respectively $\{t,r,\theta,\tilde{\varphi}_\pm\}$ to $h_\pm$.) Since we are interested only in the time-radial part of the metric, we must consider $\tilde{\varphi}_\pm =const.$ So, $d\tilde{\varphi}_\pm=0$ and therefore $$(d\varphi-{{a}\over{r^2+a^2}}dt)^2={{4a^2\alpha_\pm^2}\over{(r_\pm^2+a^2)^4}}(\rho^2)^2dt^2,\eqno{(8)}$$ which can be neglected to $O(\rho^2)$.

ii) For the $dr^2$ term: $dr=d(r_\pm+{{\alpha_\pm}\over{r_\pm}}\rho^2)={{2\alpha_\pm}\over{r_\pm}}\rho d\rho$, so $dr^2={{4\alpha_\pm^2}\over{r_\pm^2}}\rho^2d\rho^2$. The surface gravities at $r_\pm$ are given by (see Appendix) $$\kappa_+={{r_+-r_-}\over{2(r_+^2+a^2)}}={{\sqrt{M^2-(a^2+Q^2)}}\over{2M(M+\sqrt{M^2-(a^2+Q^2)})-Q^2}}\eqno{(9a)}$$ and $$\kappa_-=-{{r_+-r_-}\over{2(r_-^2+a^2)}}=-{{\sqrt{M^2-(a^2+Q^2)}}\over{2M(M-\sqrt{M^2-(a^2+Q^2)})-Q^2}}\eqno{(9b)}$$ with $\kappa_+>0$, $\kappa_-<0$, and $\vert\kappa_-\vert >\kappa_+$ since $r_-<r_+$. From (9a), $r_+-r_-=2(r_+^2+a^2)\kappa_+$, which implies $2(r_+^2+a^2)\kappa_+-r_+=-r_-$, then $r-r_-=2(r_+^2+a^2)\kappa_++(r-r_+)=2(r_+^2+a^2)\kappa_++{{\alpha_+}\over{r_+}}\rho^2$ and therefore $$\Delta=(r-r_+)(r-r_-)={{\alpha_+}\over{r_+}}(2(r_+^2+a^2)\kappa_++{{\alpha_+}\over{r_+}}\rho^2)={{2\alpha_+}\over{r_+}}(r_+^2+a^2)\kappa_+\rho^2+O(\rho^4); \eqno{(10a)}$$ analogously, from (9b), $$\Delta=(r-r_+)(r-r_-)={{\alpha_-}\over{r_-}}(2(r_-^2+a^2)\kappa_-+{{\alpha_-}\over{r_-}}\rho^2)={{2\alpha_-}\over{r_-}}(r_-^2+a^2)\kappa_-\rho^2+O(\rho^4). \eqno{(10b)}$$ On the other hand, $$\Sigma=r^2+a^2cos^2\theta=(r_\pm+{{\alpha_\pm}\over{r_\pm}}\rho^2)^2+a^2cos^2\theta=r_\pm^2+a^2cos^2\theta+2\alpha_\pm\rho^2+O(\rho^4)$$ $$:=\Sigma_\pm+2\alpha_\pm\rho^2+O(\rho^4)\eqno{(11)}$$ with $$\Sigma_\pm=r_\pm^2+a^2cos^2\theta.\eqno{(12)}$$ Then, up to $O(\rho^2)$ terms, $$-{{\Sigma}\over{\Delta}}dr^2=-{{4\alpha_\pm\Sigma_\pm}\over{r_\pm(r_\pm-r_\mp)}}d\rho^2. \eqno{(13)}$$

\

iii) For the third term in (1), at fixed $\theta$, $$-\Sigma d\theta^2=0.\eqno{(14)}$$

\

iv) For the first term in (1), from (10a) and (10b), $\Delta={{\alpha_\pm}\over{r_\pm}}(r_\pm-r_\mp)\rho^2$, and then ${{\Delta}\over{\Sigma}}={{\alpha_\pm(r_\pm-r_\mp)}\over{r_\pm\Sigma_\pm}}\rho^2$. Also, again for constant values of $\tilde{\varphi}_\pm$, $$dt-asin^2\theta d\varphi=dt-asin^2\theta(d\tilde{\varphi}+\omega_\pm dt)=dt-asin^2\theta\omega_\pm dt$$ $$=dt(1-{{a^2sin^2\theta}\over{r_\pm^2+a^2}})=dt({{\Sigma_\pm}\over{r_\pm^2+a^2}}).$$ Then, $${{\Delta}\over{\Sigma}}(dt-asin^2\theta d\varphi)^2=\Sigma_\pm{{\alpha_\pm(r_\pm-r_\mp)}\over{r_\pm(r_\pm^2+a^2)^2}}\rho^2dt^2={{4\Sigma_\pm\alpha_\pm}\over{r_\pm(r_\pm-r_\mp)}}(\kappa_\pm\rho)^2dt^2.
\eqno{(15)}$$ So, from (8), (13), (14) and (15), $$ds^2\vert_{\matrix{\theta,\tilde{\varphi}_\pm=constants\cr r\sim r_\pm\cr}}={{4\Sigma_\pm\alpha_\pm}\over{r_\pm(r_\pm-r_\mp)}}((\vert\kappa_\pm\vert\rho)^2dt^2-d\rho^2),\eqno{(16)}$$ which is conformal to the {\it Rindler metric with proper acceleration} $\vert\kappa_\pm\vert>0$ [5-6]. Choosing $$\alpha_\pm:=\tilde{\alpha}_\pm={{r_\pm(r_\pm-r_\mp)}\over{4\Sigma_{\pm}}}=\{\matrix{{{r_+(r_+-r_-)}\over{4(r_+^2+a^2cos^2\theta)}}>0 \ at \ h_+, \cr -{{r_-(r_+-r_-)}\over{4(r_-^2+a^2cos^2\theta)}}<0 \ at \ h_-,}\eqno{(17)}$$ one obtains $$ds^2\vert_{\matrix{\theta,\tilde{\varphi}_\pm=constants \cr r\sim r_\pm \cr \alpha_\pm=\tilde{\alpha}_\pm}}=ds^2_{Rindler}=(\vert\kappa_\pm\vert\rho)^2dt^2-d\rho^2\eqno{(18)}$$ with the change of coordinates in (3) given by $$r=r_\pm+{{\tilde{\alpha}_\pm(\theta)}\over{r_\pm}}\rho^2=\{\matrix{r_++{{r_+-r_-}\over{4\Sigma_+(\theta)}}\rho^2>r_+ \ near \ h_+,\cr r_--{{r_+-r_-}\over{4\Sigma_-(\theta)}}\rho^2<r_- \ near \ h_-.}\eqno{(19)}$$ Thus, the choice of the local Rindler coordinate system ($t,\rho$) depends on $\theta$. In particular, at the equator, $$\tilde{\alpha}_\pm({{\pi}\over{2}})={{r_\pm-r_\mp}\over{4r_\pm}}=\{\matrix{{{r_+-r_-}\over{4r_+}}>0 \ at \ h_+,\cr -{{r_+-r_-}\over{4r_-}}<0 \ at \ h_-.}\eqno{(20)}$$ The points at $r_+(\rho)$ near $r_+$ are outside $h_+$ but inside $S_+$, the external ergosphere, while the points at $r_-(\rho)$ near $r_-$ are inside $h_-$ but outside $S_-$, the internal ergosphere, and therefore outside the region with $-\infty<r<0$ where closed timelike curves exist. In the first two regions the metric coefficients $g_{tt}$ and $g_{rr}$ are negative; the metric, however, is Lorentzian. In Figure 1, the shadowed regions in the elementary cell of the Penrose-Carter diagram of the Kerr-Newman spacetime, indicate the zones of validity of the Rindler approximation. The total geometry in these zones is the product of the flat Rindler space and the 2-sphere.

\

Finally, the Hawking temperatures at the horizons can be read as an Unruh effect [7] due to the uniformly accelerated motions with accelerations $\vert\kappa_\pm\vert$, and are given by $$\tau_\pm={{\vert\kappa_\pm\vert}\over{2\pi}}.\eqno{(21)}$$

\begin{figure}[h!]

  \centering
    \includegraphics[width=1\textwidth]{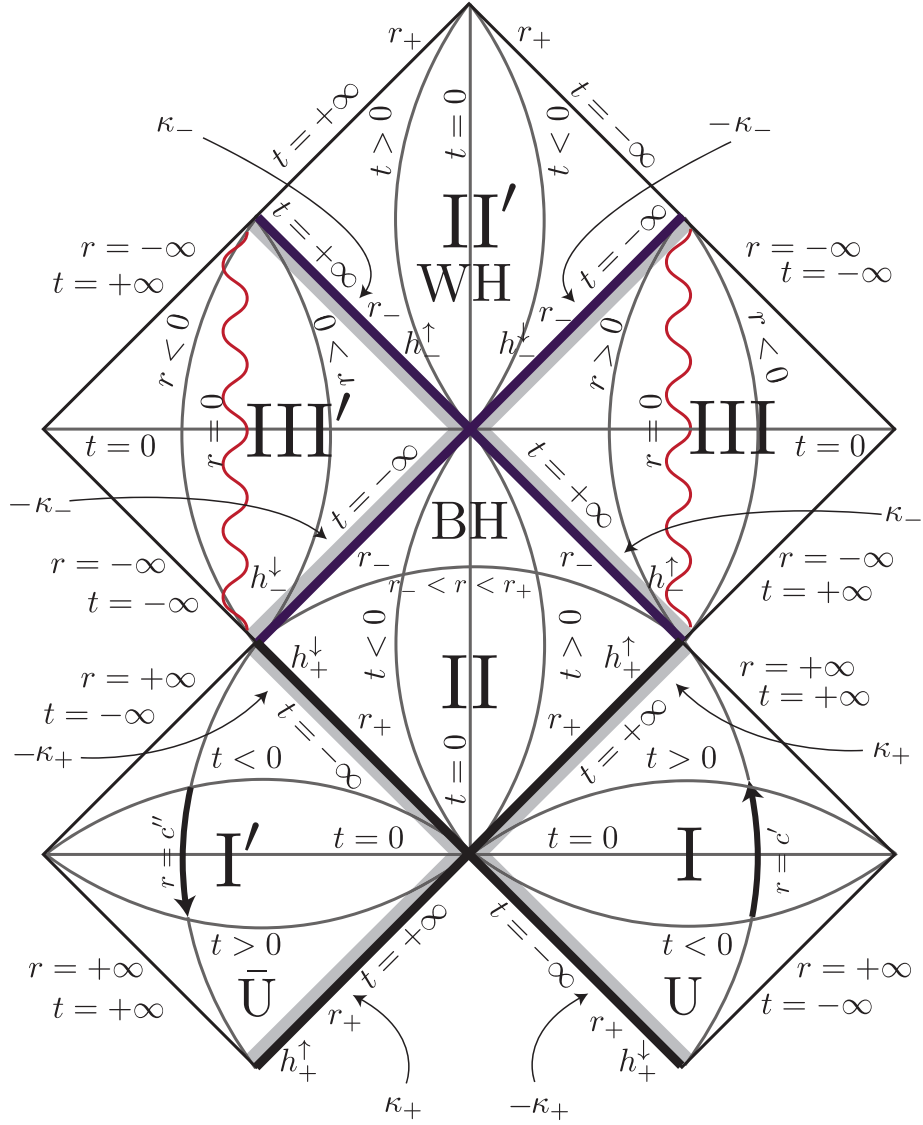}
     \caption{In the shadowed regions Rindler aproximation holds; $\pm\kappa_\pm,\mp\kappa_\pm$: surface gravities; $h^{\uparrow\downarrow}_\pm$: future ($\uparrow$) and past ($\downarrow$) event (+) and Cauchy (-) horizons; $U$: universe, $\bar{U}$: anti-universe; BH: black hole; WH: white hole.}
\end{figure}

{\bf 2 Comment on the approximation}

\

From (3) and (17), it can be easily seen that $O(\rho^4)<<O(\rho^2)$ means that $$\vert\rho\vert<<2\sqrt{{{r_\pm\Sigma_\pm}\over{r_+-r_-}}}.\eqno{(22)}$$ In particular, for $\theta={{\pi}\over{2}}$,

$$\vert\rho({{\pi}\over{2}})\vert<<2\sqrt{{{r_\pm^3}\over{r_+-r_-}}}=\{\matrix{{{2r_+}\over{\sqrt{1-{{r_-}\over{r_+}}}}} \ at \ r_+,\cr {{2r_-}\over{\sqrt{{{r_+}\over{r_-}}-1}}} \ at \ r_-.}\eqno{(23)}$$ For example, for ${{r_-}\over{r_+}}={{1}\over{2}}$ i.e. for $M^2={{9}\over{8}}(a^2+Q^2)$, $$\vert\rho({{\pi}\over{2}})\vert <<\{\matrix{2\sqrt{2}r_+ \ at \ r_+,\cr 2r_- \ at \ r_-.}\eqno{(24)}$$

\

{\bf Appendix}

\

In this Appendix we present a detailed derivation of the surface gravities $\kappa_\pm$ at the horizons $h_\pm$.

\

In the retarded Kerr coordinates $(u,r,\theta,\chi)$ (Eddington-Finkelstein type), with $u,r\in(-\infty,+\infty)$, $\theta\in [0,\pi]$, and $\chi\in [0,2\pi)$, the Kerr-Newman metric is given by $$g_{\mu\nu}=\pmatrix{g_{uu}&g_{ur}&g_{u\theta}&g_{u\chi}\cr \cdot&g_{rr}&g_{r\theta}&g_{r\chi}\cr \cdot&\cdot&g_{\theta\theta}&g_{\theta\chi}\cr \cdot&\cdot&\cdot&g_{\chi\chi}\cr}=\pmatrix{1-{{2Mr-Q^2}\over{\Sigma}}&1&0&asin^2\theta{{2Mr-Q^2}\over{\Sigma}}\cr \cdot&0&0&-asin^2\theta\cr \cdot&\cdot&-\Sigma&0\cr \cdot&\cdot&\cdot&-sin^2\theta{{A}\over{\Sigma}}\cr}\eqno{(A.1.)}$$ with inverse $$g^{\mu\nu}=\pmatrix{g^{uu}&g^{ur}&g^{u\theta}&g^{u\chi}\cr \cdot&g^{rr}&g^{r\theta}&g^{r\chi}\cr \cdot&\cdot&g^{\theta\theta}&g^{\theta \chi}\cr \cdot&\cdot&\cdot&g^{\chi\chi}\cr}=\pmatrix{{{-a^2sin^2\theta}\over{\Sigma}}&{{r^2+a^2}\over{\Sigma}}&0&-{{a}\over{\Sigma}}\cr \cdot&-{{\Sigma-2Mr+Q^2+a^2sin^2\theta}\over{\Sigma}}&0&{{a}\over{\Sigma}}\cr \cdot&\cdot&-{{1}\over{\Sigma}}&0\cr \cdot&\cdot&\cdot&-{{1}\over{\Sigma sin^2\theta}}}.\eqno{(A.2.)}$$ In (A.1.), $$A=\Sigma(r^2+a^2)+a^2sin^2\theta{{2Mr-Q^2}\over{\Sigma}}=(r^2+a^2)^2-a^2sin^2\theta\Delta.\eqno{(A.3.)}$$ Since $\partial_ug_{\mu\nu}=0$ and $\partial_\chi g_{\mu\nu}=0$, $\partial_u$ and $\partial_\chi$ are Killing vector fields.

\

The horizons $h_\pm$ are given by the equations $${\cal K}^\pm(r)=0\eqno{(A.4.)}$$ where ${\cal K}^\pm$ are the hypersurfaces defined by $${\cal K}^\pm(r)=r-r_\pm.\eqno{(A.5.)}$$ Their normal vector fields are $$l_\pm=\tilde{f}_\pm g^{\mu\nu}(\partial_\nu{\cal K}^\pm)\partial_\mu=l^u_\pm\partial_u+l^r_\pm\partial_r+l^\chi_\pm\partial_\chi\eqno{(A.6.)}$$ where $$l^u_\pm=\tilde{f}_\pm{{r^2+a^2}\over{\Sigma}}, \ l^r_\pm=-{{\tilde{f}_\pm\Delta}\over{\Sigma}}, \ l^\chi_\pm=\tilde{f}{{a}\over{\Sigma}},\eqno{(A.7)}$$ and $\tilde{f}_\pm$ are arbitrary non vanishing functions.

\

${\cal K}^\pm$ are null surfaces: in fact, an easy calculation leads to $$l^2_\pm\vert_{r_\pm}=(g_{\mu\nu}l^\mu_\pm l^\nu_\pm)\vert_{r_\pm}\sim\Delta(r_\pm)=0\eqno{(A.8.)}$$ with $$l_\pm\vert_{r_\pm}=\tilde{f}_\pm{{r_\pm^2+a^2}\over{\Sigma_\pm}}\xi_\pm\vert_{r_\pm}:=(f_\pm)^{-1}\xi_\pm\vert_{r_\pm},\eqno{(A.9)}$$ where $f_\pm={{\Sigma_\pm}\over{r_\pm^2+a^2}}(\tilde{f}_\pm)^{-1}$ and $$\xi_\pm\vert_{r_\pm}=\partial_u\vert_{r_\pm}+\omega_\pm\partial_\chi\vert_{r_\pm}, \ \xi_\pm^2\vert_{r_\pm}=0\eqno{(A.10)}$$ i.e. $$\xi^u_\pm\vert_{r_\pm}=1, \ \xi^r_\pm\vert_{r_\pm}=0, \ \xi^\chi_\pm\vert_{r_\pm}=\omega_\pm.\eqno{(A.11)}$$ Being a linear combination of Killing vectors, $\xi_\pm\vert_{r_\pm}$ are also Killing vectors. Then, $h_+$ and $h_-$ are Killing horizons.

\

For later use, from (A.7) and the definition (A.10), $$\xi^u_\pm=1, \ \xi^r_\pm=-{{\Delta}\over{r^2+a^2}}, \ \xi^\chi_\pm={{a}\over{r^2+a^2}}.\eqno{(A.12.)}$$

\

If $t^\mu_\pm$ are tangent to ${\cal K}_\pm$, then $t_\pm\cdot l_\pm\vert_{r_\pm}=0$, and since $l^2_\pm\vert_{r_\pm}=0$, then $l_\pm\vert_{r_\pm}$ are also tangent to ${\cal K}_\pm$. So, $l^\mu_\pm\vert_{r_\pm}={{dx^\mu}\over{d\lambda}}$ for some curve $x^\mu(\lambda)\subset{\cal K}_\pm.$ If $\lambda$ is an affine parameter, then [8] $$l\cdot Dl^\mu_\pm\vert_{r_\pm}=0\eqno{(A.13)}$$ and the Killing vectors obey the equation $$\xi_\pm\cdot D\xi_\pm\vert_{r_\pm}=\kappa_\pm\xi_\pm\vert_{r_\pm}\eqno{(A.14)}$$ where $\kappa_\pm=\xi_\pm\cdot\partial(ln\vert f_\pm\vert)$. ($D$ and $\partial$ are respectively covariant and ordinary derivatives.) The l.h.s. of (A.13) can be written $$\xi_\pm\cdot D\xi_{\pm\mu}=-{{1}\over{2}}\xi^2_{\pm,\nu}\eqno{(A.15)}$$ which from $$-{{1}\over{2}}\xi^2_{\pm,\nu}\vert_{r_\pm}=\kappa_\pm(\xi_\pm)_\nu\vert_{r_\pm},\eqno{(A.16)}$$ allows the determination of $\kappa_\pm$.

\

From (A.1) and (A.11) one obtains $$(\xi_\pm)_u\vert_{r_\pm}=(\xi_\pm)_\theta\vert_{r_\pm}=(\xi_\pm)_\chi\vert_{r_\pm}=0, \ (\xi_\pm)_r\vert_{r_\pm}=1-a\omega_\pm sin^2\theta.\eqno{(A.17)}$$ So, from (A.15), $$-{{1}\over{2}}\xi^2_{\pm,r}\vert_{r_\pm}=\kappa_\pm(\xi_\pm)_r\vert_{r_\pm}=(1-a\omega_\pm sin^2\theta)\kappa_\pm={{\Sigma}\over{r^2_\pm+a^2}}\kappa_\pm.\eqno{(A.18)}$$ Using (A.12), $$\partial_r\xi^2_\pm\vert_{r_\pm}=\partial_r(\xi_\pm)_u\vert_{r_\pm}-{{1}\over{r^2_\pm+a^2}}\partial_r\Delta\vert_{r_\pm}(\xi_\pm)_r\vert_{r_\pm}+{{a}\over{r^2_\pm+a^2}}\partial_r(\xi_\pm)_\chi\vert_{r_\pm},$$ $$\partial_r\Delta\vert_{r_\pm}=2(r_\pm-M)=r_\pm-r_\mp, \eqno{(A.19)}$$ and from $$(\xi_\pm)_u=g_{u\mu}\xi_\pm^\mu, \ and \ (\xi_\pm)_\chi=g_{\chi\mu}\xi_\pm^\mu, \eqno{(A.20)}$$ a long but straightforward calculation gives $$\partial_r(\xi_\pm)_u\vert_{r_\pm}=0, \ and \ \partial_r(\xi_\pm)_\chi\vert_{r_\pm}=0.\eqno{(A.21)}$$ Putting these results together in (A.18) one finally obtains $$\kappa_\pm={{r_\pm-r_\mp}\over{2(r_\pm^2+a^2)}}.\eqno{(A.22)}$$

\

{\bf Acknowledgments}

\

This work was partially supported by the project PAPIIT IN105413, DGAPA, UNAM. Figure 1 was done by O. Brauer.

\

\section*{References}

1. Kerr, R. P.: Gravitational Field of a Spinning Mass as an Example of Algebraically Special Metrics. Phys. Rev. Lett. {\bf 11}, 237-238 (1963)

\

2. Newman, E. T., Couch, E., Chinnapared, K., Exton, A., Prakash, A., and Torrence, R.: Metric of a Rotating, Charged Mass. J. Math. Phys. {\bf 6}, 918-919 (1965)

\

3. Boyer, R. H. and Lindquist, R. W.: Maximal Analytic Extension of the Kerr Metric. J. Math. Phys. {\bf 8}, 265-281 (1967)

\

4. Cheng, T. P.: Relativity, Gravitation and Cosmology. A Basic Introduction. Oxford University Press, New York (2010)

\

5. Rindler, W.: Kruskal Space and the Uniformly Accelerated Frame. Am. J. Phys. {\bf 34}, 1174-1178 (1966)

\

6. 'tHooft, G.: Introduction to the Theory of Black Holes. Lecture Notes, Utrecht University (2009)

\

7. Unruh, W. G.: Notes on black holes evaporation. Phys. Rev. D {\bf 14}, 870-892 (1976)

\

8. Townsend, P. K.: Black Holes. Lecture Notes. DAMTP, University of Cambridge (1997); arXiv: gr-qc/9707012v1

\end{document}